\documentstyle[12pt]{article}

\begin{document}

\begin{flushright}
IMSc/2006/03/08 
\end{flushright} 

\vspace{2mm}

\vspace{2ex}

\begin{center}
{\large \bf A Stringy Correspondence Principle in Cosmology } \\ 



\vspace{8ex}

{\large  S. Kalyana Rama}

\vspace{3ex}

Institute of Mathematical Sciences, C. I. T. Campus, 

Tharamani, CHENNAI 600 113, India. 

\vspace{1ex}

email: krama@imsc.res.in \\ 

\end{center}

\vspace{6ex}

\centerline{ABSTRACT}
\begin{quote} 

We study a $d-$dimensional FRW universe, containing a perfect
fluid with $p = w \rho$ and $\frac{1} {d - 1} \le w \le 1$, and
find a correspondence principle similar to that of Horowitz and
Polchinski in the black hole case. This principle follows quite
generally from thermodynamics and the conservation of energy
momentum tensor, and can be stated along similar lines as in the
black hole case: ``When the temperature $T$ of the universe
becomes of order string scale the universe state becomes a
highly excited string state. At the transition, the entropies
and energies of the universe and strings differ by factors of
${\cal O}(1)$.'' Such a matching is absent for $w \ne 1$ if the
transition is assumed to be when the curvature or the horizon
length is of order string scale.

\end{quote}

\vspace{2ex}


\newpage

\vspace{4ex}

{\bf 1.}  
Consider a Schwarzschild black hole. As it evaporates through
Hawking radiation, its mass and radius $r_{Sch}$ decrease and
its temperature $T_{bh}$ increases, all eventually reaching
Planckian scales at which Einstein's equations are no longer
valid.

In string theory, gravity arises as a low energy mode and
Einstein's equations are obtained as a low energy approximation.
Planck length $l_{pl}$ is related related to the string length
$l_s$ and the string coupling constant $g_s$, taken to be $\le
1$. Given this origin of gravity, it is natural to expect that
when $r_{Sch} \simeq l_s$ a stringy description will take over.

This idea was proposed by Susskind in \cite{s} and later
incorporated by Horowitz and Polchinski in their correspondence
principle for black holes and strings \cite{hp}. This principle
states that when $r_{Sch} \simeq l_s$ a black hole state becomes
a string state. At this transition, masses and entropies of
black hole and strings differ by numerical factors of ${\cal
O}(1)$ and not by, say, $g_s-$dependent factors which can be
$\gg 1$ or $\ll 1$. This is true for a variety of black holes
\cite{hp}.

In cosmology also, as time decreases in the past, the
temperature generically increases eventually reaching Planckian
scale. It is then natural to wonder whether a similar
correspondence principle exists in cosmology. In the FRW
universe we study here, we find that such a principle indeed
exists and follows quite generally from thermodynamics and the
conservation of energy momentum tensor. It is applicable near
big bang/crunch singularities and is likely to be applicable in
more general cases also wherever the temperature is expected to
reach string scale.

This correspondence principle in cosmology can be stated along
similar lines as in the black hole case: When the temperature
$T$ of the universe becomes of the order of string scale the
universe state becomes a string state containing highly excited
strings. At the transition, the entropies and energies of the
universe and strings differ by factors of ${\cal O}(1)$. These
factors presumably depend on when the transition takes place and
cannot be calculated within the present approach.

The criterion $T \simeq \frac{1} {l_s}$ is equivalent to $S
\simeq S_s$ where $S$ is the entropy of the universe contained
within its horizon and $S_s$ is the entropy of the strings with
the same amount of energy as in the universe within its horizon.
It turns out that this is the correct criterion for the
transition and not, for example, $L_H \simeq l_s$ or ${\cal R}
\simeq \frac{1}{l_s^2}$ where $L_H$ is the physical size of the
horizon and ${\cal R}$ is the curvature scale. If the later
criteria are used then, at the transition, the entropies and
energies differ by $g_s-$dependent factors. Also, the criterion
$T \simeq \frac{1} {l_s}$ or, equivalently, $S \simeq S_s$ is
more general since, applied to black holes using suitably red
shifted black hole temperatures, it yields the same results as
in \cite{hp}.

In this paper, we consider the evolution of a $d-$dimensional
spatially flat homegeneous isotropic universe containing a
perfect fluid with density $\rho$ and pressure $p = w \rho$. We
assume that $d \ge 4$ and, since high temperatures are involved,
that $\frac{1}{d - 1} \le w \le 1$. We also assume that the
universe saturates the Fischler-Susskind holographic bound
\cite{fs} at Planckian times in the beginning. This ensures that
numerical coefficients in various expressions are all of ${\cal
O}(1)$.

In section {\bf 2} we show that for the above FRW universe, a
transition to string state is likely to occur at $T \simeq
\frac{1} {l_s}$. In section {\bf 3} we show that the main
results follow quite generally from thermodynamics and the
conservation of energy momentum tensor. In section {\bf 4} we
explain this transition intuitively in a few different ways,
similar to the black hole case. At the end of this section and
in section {\bf 5} we mention some issues that may be studied
further.

\vspace{4ex}

{\bf 2.}  
Consider the evolution of a $d-$dimensional spatially flat
homegeneous isotropic universe containing a perfect fluid with
density $\rho$ and pressure $p = w \rho$. We assume that $d \ge
4$ and, since high temperatures are involved, that $\frac{1}
{d - 1} \le w \le 1$. The relevent line element $d s$ is given
by
\begin{equation}\label{dsd}
d s^2 = - d t^2 + a^2 \left( d r^2 + r^2 d \Omega_{d - 2}^2
\right)
\end{equation}
where $a(t)$ is the scale factor and $d \Omega_{d - 2}$ is the
line element on an unit $(d - 2)-$dimensional sphere. Einstein's
equations are
\begin{eqnarray}
(d - 1) (d - 2) h^2 & = & 2 \kappa^2 \rho \nonumber \\
\rho_t + (d - 1) (1 + w) h \rho & = & 0  \label{h2d}  
\end{eqnarray}
where the suffix $t$ denotes time derivative. In Planck units
defined by
\begin{equation}\label{plunits}
\hbar = c = 1 \; \; , \; \; \; 
2 \kappa^2 = l_{pl}^{d - 2} = t_{pl}^{d - 2} \; , 
\end{equation}
the solution to the above equations can be written as
\begin{equation}
a(t) = a_{pl} \left( \frac{t}{t_{pl}} \right)^\alpha
\; \; , \; \; \; \; 
\rho(t) = \rho_{pl} \left( \frac{t}{t_{pl}} \right)^{- 2} 
= \frac{\alpha^2 (d - 1) (d - 2)}{l_{pl}^{d - 2} \; \; t^2} 
\end{equation} 
where $\alpha = \frac{2}{(d - 1) (1 + w)}$. Equivalently, $a(t)$
and $\rho(t)$ can be expressed in terms of $a(t_0)$ and
$\rho(t_0)$ where $t_0$ is some conveniently chosen initial
time. The relations between $(a(t_0), \rho(t_0))$ and $(a_{pl},
\rho_{pl})$ are easy to obtain.

During the cosmological evolution the comoving entropy density
$\sigma$, given by
\[
\sigma = \frac{(\rho + p)}{T} \;  a^{d - 1}
\]
where $T$ is the temperature, remains constant. We assume that
this constant is such that, at $t = t_{pl}$, it saturates the
holographic bound \cite{fs}
\begin{equation}
\sigma \; V_{d - 1} r_H^{d - 1} \le 
\frac{C \; \omega_{d - 2} L_H^{d - 2}}{l_{pl}^{d - 2}}
\end{equation} 
where $C$ is a constant of ${\cal O}(1)$, $V_n$ is the volume of
an unit $n-$dimensional ball, $\omega_{n - 1}= n V_n =
\frac{2 \pi^{\frac{n}{2}}} {\Gamma \left( \frac{n}{2} \right)}$
is the area of an unit $(n - 1)-$dimensional sphere, and
\[
r_H = \int_0^t \frac{d t}{a} = \frac{t_{pl}}{(1 - \alpha) a_{pl}} \; 
\left( \frac{t}{t_{pl}} \right)^{1 - \alpha} 
\; \; , \; \; \; 
L_H = r_H a = \frac{t}{1 - \alpha} 
\]
are, respectively, the coordinate and the physical size of the
horizon. For more details on and a precise formulation of the
holographic bound, see \cite{fs, bousso}. It then follows that
\begin{eqnarray}
\sigma & = & C (d - 1) (1 - \alpha) \; \left( \frac{a_{pl}}{l_{pl}}
\right)^{d - 1} \\
T & = & \frac{\alpha^2 (d - 2) (1 + w)}
{C (1 - \alpha) \; \; l_{pl}} \left( \frac{t}{t_{pl}} 
\right)^{- \frac{2 w}{1 + w}} \; .
\end{eqnarray}
The entropy $S$ and the energy $E$, contained within a region
whose comoving coordinate is $r$ and the physical size is 
$L = r a$, are
\begin{equation}\label{ser}
S = \sigma \; V_{d - 1} r^{d - 1} 
\; \; , \; \; \; 
E = \rho \; V_{d - 1} L^{d - 1} \; . 
\end{equation}

In the past the time decreases and the temperature increases and
eventually reaches Planckian scale at which Einstein's equations
are no longer valid. In string theory, gravity arises as a low
energy mode and Einstein's equations are obtained as a low
energy approximation. Planck length $l_{pl}$ can be taken to be
related to the string length $l_s$ and the string coupling
constant $g_s$, taken to be $\le 1$, as follows:
\begin{equation}\label{pltos}
l^{d - 2}_{pl} = g_s^2 \; l_s^{d - 2} 
\; \; \; \longleftrightarrow \; \; \; 
l_{pl} = g_s^X \; l_s \; \; , \; \; \; 
X = \frac{2}{d - 2} \; .
\end{equation}
In the past, as time decreases, the temperature will increase
and first reach the string scale $\simeq \frac{1}{l_s}$. At such
a scale higher modes of strings will be excited copiously,
because high energies are involved, and the universe state
becomes a string state containing highly excited strings. The
FRW description of the universe, given above, is then replaced
by a stringy description. 

That such a transition is likely can be seen from comparing the
entropy $S$ of the universe with the entropy $S_s$ of the
strings with the same amount of energy. The entropy $S$ and the
energy $E$ of the universe depend on its size. It is natural to
take this size to be given by the horizon with comoving
coordinate $r_H$ and the physical size $L_H = r_H a$ as given
above. This is the maximum size within which causal contact is
possible. In the following, we do so and consequently $S$ and
$E$ will refer to the entropy and the energy of the universe
contained within its horizon unless mentioned otherwise. From
the equations given above, it then follows that
\begin{eqnarray} 
\rho & = & \frac{C_\rho}{l_{pl}^{d - 2} \; t^2} 
\nonumber \\
T & = & \frac{C_T}{l_{pl}} \left( \frac{t}{t_{pl}} 
\right)^{- \frac{2 w}{1 + w}}
\nonumber \\
S & = & C_S \left( \frac{t}{t_{pl}} 
\right)^{(d - 1) (1 - \alpha)} 
\nonumber \\
E & = & \frac{C_E}{l_{pl}} \left( \frac{t}{t_{pl}} 
\right)^{d - 3}
\end{eqnarray}
where the constants $C_\rho$, $C_T$, $C_S$, and $C_E$ are all of 
${\cal O}(1)$ and are given by 
\begin{eqnarray}
C_\rho & = & \alpha^2 (d - 1) (d - 2)
\; \; , \; \; \; 
C_T = \frac{\alpha^2 (d - 2) (1 + w)}{C (1 - \alpha)}
\nonumber \\
C_E & = & C_\rho V_{d - 1}
(1 - \alpha)^{- (d - 1)} 
\; \; , \; \; \; 
C_S = C \omega_{d - 2} (1 - \alpha)^{- (d - 2)} \; .
\end{eqnarray}
The above expressions for $(\rho, T, S, E)$ can also be
expressed in terms of $g_s$ and $l_s$ using $l_{pl} = g^X l_s$,
see equation (\ref{pltos}).

The entropy $S_s$ of highly excited strings having energy $E$
is given by
\begin{equation}\label{ss}
S_s (E) = C_s l_s E
\end{equation}
where $C_s$ is a numerical constant of ${\cal O}(1)$. The ratio
of the entropy $S$ of the universe to the entropy $S_s$ of
highly excited strings with the same energy can now be
calculated and is given by
\begin{equation}
\frac{S}{S_s} = \frac{1 + w}{C_s l_s T} 
= \frac{T_*}{T}
\; \; \;, \; \; \; \; 
T_* = \frac{1 + w}{C_s l_s}
\end{equation}
where the temperature $T_*$ is of the order of string scale.
This shows that if the temperature of the universe $T > T_*$
then its entropy $S < S_s$ and, hence, the stringy phase is
entropically favourable and the state becomes a string state. At
lower temperatures $S > S_s$ and, hence, the FRW phase is
entropically favourable and the state becomes a FRW state. At
the transition, we have
\[
T \simeq \frac{1}{l_s}
\; \; , \; \; \; 
S \simeq \left( g_s^X l_s E \right)^{\frac{(d - 1) (1 - \alpha)}
{d - 3}} \simeq S_s \simeq l_s E \simeq \sqrt{N}
\] 
where $N \gg 1$ is the string excitation number and, here and in
the following, $A \simeq B$ means that $A$ and $B$ are equal
upto numerical factors of ${\cal O}(1)$. It also follows that
the string coupling constant \footnote{We used $X = \frac{2}{d -
2}$ and the identities $\frac{(d - 1) (1 - \alpha)}{d - 3} =
\frac{d - 3 + (d - 1) w}{(d - 3) (1 + w)} = 1 + \frac{2 w}{(d -
3) (1 + w)}$.}
\begin{equation}\label{gamma}
g_s \simeq N^{- \gamma} \ll 1 
\; \; , \; \; \; 
\gamma = \frac{(d - 2) w}{2 (d - 3 + (d - 1) w)}
\end{equation}
since $\frac{1}{2 (d - 1)} \le \gamma \le \frac{1}{4}$ when $w$
lies in the range $\frac{1}{d - 1} \le w \le 1$ as assumed here.
Since $g_s \ll 1$ it is consistent to use the weak coupling
formulas for strings. Also, note that the transition temperature
in Planck units $l_{pl} T_* \simeq g_s^X \ll 1$ so that the
universe is well away from Planckian regime.

This suggests the following correspondence between a FRW
universe state and a highly excited string state: When the
temperature is lower than string scale, the universe state
evolves as in FRW cosmology. When the temperature becomes of the
order of string scale the universe state becomes a string state
containing highly excited strings. At the transition, the
entropies and energies of the universe and strings differ by
factors of ${\cal O}(1)$. These factors presumably depend on
when the transition takes place and cannot be calculated within
the present approach.

\vspace{4ex}

{\bf 3.}  
The main results above follow quite generally and are valid for
spatially curved homogeneous isotropic universes also. We have,
from thermodynamics, that $p_T = \frac{\rho + p}{T}$ \cite{w}
which, together with energy momentum conservation equation
$\rho_t + (d - 1) h (\rho + p) = 0$, implies that the comoving
entropy density $\sigma = \left(\frac{\rho + p}{T}\right) \;
a^{d - 1} = constant$. Let $S$ and $E$ be the entropy and the
energy of the universe contained within a region whose comoving
coordinate is $r$ and the physical size is $L = r a$.  Then, it
follows that
\[
S = \sigma V_{d - 1} r^{d - 1} \; \; , \;\; \; 
E = \rho V_{d - 1} L^{d - 1} 
\; \; , \;\; \; 
\frac{S}{E} = \frac{\rho + p}{\rho \; T} \; .
\]
If one assumes that $p = w \rho$ then it follows that $S =
\frac{(1 + w) E}{T}$.

The ratio of the entropy $S$ of the universe to the entropy
$S_{s}$, given in equation (\ref{ss}), of the highly excited
strings with the same energy is thus given by
\begin{equation}
\frac{S}{S_s} = \frac{1 + w}{C_s l_s \; T} 
= \frac{T_*}{T}
\; \; \; , \; \; \; \; 
T_* = \frac{1 + w}{C_s l_s}
\end{equation}
where the temperature $T_*$ is of the order of string scale.
This shows, as before and more generally, that if the
temperature of the universe $T > T_*$ then its entropy $S < S_s$
and, hence, the stringy phase is entropically favourable and the
state becomes a string state. At lower temperatures $S > S_s$
and, hence, the FRW phase is entropically favourable and the
state becomes a FRW state. At the transition, we have
\[
T \simeq \frac{1}{l_s}
\; \; , \; \; \; 
S \simeq l_s E \simeq S_s \simeq \sqrt{N} 
\] 
where $N \gg 1$ is the string excitation number. 

In particular, the above results are valid near a big crunch
singularity also which occurs in future in a closed universe and
where the temperature is expected to reach string scale. One
then infers that a FRW universe state becomes a string state
near a big crunch singularity also when the temperature $T
\simeq \frac{1}{l_s}$.

For $p = w \rho$ it follows that $T = \frac{C_T}{l_{pl}} \;
\left( \frac{a}{a_{pl}} \right)^{- (d - 1) w}$. The density
$\rho$ and the curvature scale ${\cal R}$ \footnote{The
curvature scale ${\cal R}$ is given, for example, by the Ricci
scalar $R^\mu_\mu$ (which vanishes if $T_{\mu \nu}$ is
traceless) or $\sqrt{R^{\mu \nu} R_{\mu \nu}}$.  Einstein's
equations imply that ${\cal R} \simeq \kappa^2 \rho$.} can now
be expressed as
\[
\rho = \frac{C_\rho}{l_{pl}^d} \; 
\left( \frac{l _{pl} T}{C_T} \right)^{\frac{1 + w}{w}}
\; \; , \; \; \; 
{\cal R} = \frac{C_R}{l_{pl}^2} \; 
\left( \frac{l _{pl} T}{C_T} \right)^{\frac{1 + w}{w}} 
\simeq \; \; l_{pl}^{d - 2} \; \rho 
\]
where the $C's$ are numerical constants of ${\cal O}(1)$ as will
follow from the holographic bound. The explicit time dependence 
of the various quantities can only be obtained by solving
Einstein's equations. One can then express the entropy $S(T)$
within the horizon of the universe as a function of its
temperature $T$, and thereby obtain the $N-$dependence of $g_s$
at the transition using $S\left( \frac{1}{l_s} \right) \simeq
S_s \simeq \sqrt{N}$.

At the transition, $T \simeq
\frac{1}{l_s}$ and we have
\[
\rho_* \simeq \frac{1}{g_s^b \; l_s^d} 
\; \; , \; \; \; 
{\cal R}_* \simeq \frac{g_s^{b_R}}{l_s^2} 
\]
where $b = \left(d - 1 - \frac{1}{w} \right) X$, $b_R = 2 - b =
\left(\frac{1 - w}{w} \right) X$, and $X = \frac{2}{d - 2}$.
For the range $\frac{1}{d - 1} \le w \le 1$ for $w$ assumed
here, we have
\[
\frac{1}{l_s^d} \le \rho_* \le \frac{1}{g_s^2 \; l_s^d}
\; \; , \; \; \; 
\frac{g_s^2}{l_s^2} \le {\cal R}_* \le \frac{1}{l_s^2} \; . 
\]

For weak coupling, $g_s \ll 1$. Then, at the transition, the
density is of order string scale or higher whereas the curvature
is much smaller than the string scale for $w < 1$ and is of
string scale only for $w = 1$. Still, however, the transition to
highly excited string state is likely to occur at $T \simeq
\frac{1} {l_s}$ because: at the transition, the temperature is
of order string scale; the energy $E \gg \frac{1} {l_s}$, so
higher stringy modes will be excited copiously; the stringy
phase is entropically more favourable and the state becomes a
string state.

It is unlikely, for $w < 1$, that `curvature, or the physical
horizon size, be of order string scale' is the correct criterion
for a FRW universe state to become a string state. To see this,
assume that the transition occurs when ${\cal R} \simeq \frac{1}
{l_s^2}$.  Then $g_s^{(w - 1) X} \simeq (l_s T)^{1 + w}$ at the
transition and the ratio $\frac{S} {S_s} \simeq \frac{1}{l_s T}
\simeq g^{\left( \frac{1 - w}{1 + w} \right) \; X}$ would be
$\ll 1$ at weak coupling, and not of ${\cal O}(1)$ as above. For
the spatially flat case, note that ${\cal R} \simeq \frac{1}
{L_H^2}$ and, hence, the two criteria ${\cal R} \simeq \frac{1}
{l_s^2}$ and $L_H \simeq l_s$ are equivalent to each other.

\vspace{4ex}

{\bf 4.}  
Horowitz and Polchinski have formulated a correspondence
principle for black holes where there is a transition from a
black hole state to a string state containing highly excited
strings \cite{hp} which happens when the horizon radius of the
black hole $r_{bh} \simeq l_s$. This transition can be explained
intuitively in a few different ways \cite{s, hp, s2, hprev}. The
transition from a FRW universe state to a string state can also
be explained intuitively in similar ways. For example:

\vspace{2ex}

\noindent
{\bf (i)} 
At the transition one would like to equate the energy $E$ 
of the universe within its horizon
\[
E \simeq \frac{1}{l_{pl}} \left( l_{pl} 
T \right)^{- \frac{(d - 3) (1 + w)}{2 w}} 
\]
to the string energy $E_s \simeq \frac{\sqrt{N}}{l_s}$. This
cannot be true for all values of $g_s$ since $E_s$ is
independent of $g_s$ (for weak coupling) and $E$ depends on
$g_s$ and $l_s$ through the combination $l_{pl} = g_s^X l_s$.
Therefore, the equality may hold only at one value of $g_s$
which we take to be that where $T \simeq \frac{1}{l_s}$.
Equating $E$ and $E_s$ at this value of $T$ gives $g_s \simeq
N^{- \gamma}$ with $\gamma$ given in (\ref{gamma}). It then
follows that for this value of $g_s$, the entropy $S$ of the
universe within its horizon
\[
S \simeq \left( l_{pl} T \right)^{- \frac{(d - 1) (1 - \alpha)
(1 + w)}{2 w}} \simeq \sqrt{N}
\]
becomes of the order of string entropy $S_s$. It is easy to show
that if the transition criterion had been $L_H \simeq l_s$ or
${\cal R} \simeq \frac{1}{l_s^2}$ then $S$ and $S_s$ would
differ by a $g_s-$dependent factor, which would not be of ${\cal
O}(1)$ and would not match as above. 

\vspace{2ex}

\noindent
{\bf (ii)} 
Or, consider following the state of the FRW universe as $g_s$
decreases. This means that the entropy $S$ of the universe
within its horizon remains constant. Then at the transition,
taken to occur when $T \simeq \frac{1}{l_s}$, we have
\[
g_s^X \simeq S^{- \frac{2 w}{(d - 1) (1 - \alpha) (1 + w)}} \; .
\] 
Using this value for $g_s$ and the identity $\frac{d - 3}{(d -
1) (1 - \alpha)} + \frac{2 w}{(d - 1) (1 - \alpha) (1 + w)} =
1$, it follows that the energy $E$ of the universe within its
horizon
\[
E \simeq \frac{1}{l_{pl}} \; S^{\frac{d - 3}{(d - 1) 
(1 - \alpha)}}\simeq \frac{S}{l_s}
\]
becomes of the order of string mass $E_s$. It is again easy to
show that if the transition criterion had been $L_H \simeq l_s$
or ${\cal R} \simeq \frac{1}{l_s^2}$ then $E$ and $E_s$ would
differ by a $g_s-$dependent factor, which would not be of ${\cal
O}(1)$ and would not match as above.

\vspace{2ex}

\noindent
{\bf (iii)} 
Or, consider the transition in Planck units with $l_{pl}$
fixed. Consider the universe at a temperature $T$, for example,
$\simeq 1 \; GeV$. Its relation to entropy $S$ and energy $E$ of
the universe can then be read off from the expressions given in
section {\bf 2}. Now, as $g_s$ decreases with $l_{pl}$ held
fixed, $l_s$ increases and the string temperature $\simeq
\frac{1}{l_s}$ decreases. For a Schwarzschild black hole, the
string length eventually approaches $r_{Sch} \simeq \left(
l_{pl}^{d - 2} \; M_{bh} \right)^{\frac{1}{d - 3}}$ after which
a string state description must take over. Similarly, for the
universe, the string temperature eventually approaches $T \simeq
\frac{1} {l_{pl}} \left( l_{pl} E \right)^{- \frac{2 w} {(d - 3)
(1 + w)}} \simeq 1 \; GeV$, after which a string state
description must take over.

\vspace{2ex}

In fact, the transitions from a black hole state or a FRW
universe state to a corresponding string state can both be
explained quite similarly if one makes the following
identifications:
\begin{eqnarray*}
\left( M_{bh} \; , S_{bh} \; , r_{bh} \right)
& \longleftrightarrow &
\left( E \; , S \; , T \right) \\
r_{bh} \simeq l_s 
& \longleftrightarrow &
T \simeq \frac{1}{l_s} 
\; \; \; \; \; \; or \; \; \; \; \; \; 
S(E) \simeq S_s(E) \\
r_{bh} > l_s
& \longleftrightarrow &
T < \frac{1}{l_s} \; . 
\end{eqnarray*}
The meaning of various symbols above are clear. The criterion
for transition in the black hole case is $r_{bh} \simeq l_s$; in
the FRW case it is $T \simeq \frac{1}{l_s}$ or, equivalently,
$S(E) \simeq S_s(E)$. A black hole state description is valid
for $r_{bh} > l_s$ and similarly a FRW state description is
valid for $T < \frac{1}{l_s}$; a string state description takes
over after the transition.

The criterion $T \simeq \frac{1}{l_s}$ or, equivalently, $S(E)
\simeq S_s(E)$ is more general since, applied to black holes
using suitably red shifted black hole temperatures, it yields
the same results as in \cite{hp}. This can be seen easily for
Schwarzschild black holes and can be inferred for other cases
studied in \cite{hp} because the suitably red shifted
temperatures at the transitions all turn out to be $\simeq
\frac{1}{l_s}$, as shown in \cite{hp}.

Note that in the FRW case, for weak coupling, the curvature
scale at the transition is typically less than the string
scale. One would therefore expect no higher derivative
corrections to the low energy effective action of the string
theory and, hence, to the FRW evolution dictated by it. But, at
the transition, the temperature $T \simeq \frac{1}{l_s}$ and
stringy higher modes are excited copiously because the energy at
the transition $ E_* \simeq \frac{1}{l_s} \; g_s^{- \frac{1}{2
\gamma}} \gg 1$ for weak coupling. This means that the original
low energy effective action is not valid because the energy is
not low any more and must include a huge number, of the order of
$N \simeq l_s ^2 E_*^2 \gg 1$, of fields corresponding to the
higher excitations.

It is quite interesting to ask how the reverse transition
proceeds, namely what happens when one starts from a highly
excited string state at weak coupling and increases the coupling
constant $g_s$. The string to black hole transition is studied
in detail in \cite{hp2}, see also \cite{rk}. The picture that
emerges is that as $g_s$ increases the gravitational effects
become more important and eventually the excited string state
collapses to a black hole state. 

But the present results suggest the possibility that increased
effects of gravitation may instead make the excited string state
expand into a FRW universe state. It is not clear to us what
aspects of the excited strings will decide the course of
evolution.  We note that the density $\rho_*$ at the FRW
universe to string transition has a range $ \frac{1}{l_s^d} \le
\rho_* \le \frac{1}{g_s^2 \; l_s^d} $ for $\frac{1}{d - 1} \le w
\le 1$; whereas the density $\rho_{bh*}$ at the black hole to
string transition is $ \rho_{bh*} \simeq \left( \frac{1}
{l_{pl}^{d - 2} r_{Sch}^2} \; \right)_{r_{Sch} \simeq l_s}
\simeq \frac{1}{g_s^2 \; l_s^d} \;$. This suggests that the
density of the excited strings will perhaps play a crucial role
in deciding whether, as $g_s$ increases, the excited string
state will eventually collapse to a black hole state or expand
into a FRW universe state.

Also note that at the transition $g_s \simeq N^{- \frac{1}{4}}$
for $w = 1$ case (and also in the black hole case) whereas $g_s
\simeq N^{- \frac{1}{2 (d - 1)}} > N^{- \frac{1}{4}}$ for $w =
\frac{1} {d - 1}$ case. This suggests that, as $g_s$ is
increased from $0_+$, a highly excited string state will first
expand into a $w = 1$ FRW universe (or collapse to a black
hole). As $g_s$ is increased further, it will expand into a $w =
\frac{1}{d - 1}$ FRW universe.

In this context, note that there are a lot of similarities
between the $w = 1$ case and the black hole case. Banks and
Fischler have studied them extensively \cite{bf} in connection
with, among others, black holes and early universe cosmology.
They have presented an interesting scenario where the universe
starts off with $w = 1$ matter; then black holes are nucleated
which percolate the space and eventually evaporate through
Hawking radiation; the universe thus becomes dominated by
radiation ($w = \frac{1} {d - 1}$). This sequence of evolution
is quite similar to that indicated above where one starts from a
highly excited string state and increases the string coupling.
Although the connections between these two situations are not
clear, the ideas in \cite{bf} seem very relevent to
understanding the string to black hole and/or FRW universe
transition as $g_s$ is increased.

\vspace{4ex}

{\bf 5.}  
We studied here the correspondence between a FRW universe state
to a string state for the case of isotropic homogeneous universe
containing a perfect fluid with density $\rho$ and pressure $p =
w \rho$. We assumed that $\frac{1}{d - 1} \le w \le 1$ since
high temperatures are involved. The results can also be seen to
follow from general considerations involving thermodynamics and
the conservation of energy momentum tensor.

An important consequence of this correspondence between a FRW
universe state to a string state is that the problem of big
bang/crunch singularity is obviated. This is because such
singularities are expected to occur at Planckian temperature
$T_{pl}$, whereas the transition to a string state description
takes place at a temperature which is $\ll T_{pl}$ in the weak
coupling limit.

Therefore, it is of interest to establish the generality of this
correspondence. For example one may study this correspondence for
anisotropic cases, for an universe containing more than one
perfect fluid, or containing matter whose equation of state
$p(\rho)$ is more general than $p = w \rho$ with the constant
$w$ in the range $\frac{1}{d - 1} \le w \le 1$ as assumed here.

Another interesting study is that of the transition from a
string state to black hole/FRW universe state, mentioned in more
detail in section {\bf 4}. Also, in the case of black holes,
extremal and near extremal black holes of various kinds have
provided an ideal setting to understand the black hole
$\longleftrightarrow$ string correspondence. Similar examples in
the context of cosmology will help us better understand the FRW
universe $\longleftrightarrow$ string correspondence.




\newpage

\begin{thebibliography}{999}

\bibitem{s} 
L.~Susskind,
``String theory and the principles of black hole complementarity,''
Phys.\ Rev.\ Lett.\  {\bf 71} (1993) 2367, 
arXiv: hep-th/9307168; 
\\
``Strings, black holes and Lorentz contraction,''
Phys.\ Rev.\ D {\bf 49} (1994) 6606, 
arXiv: hep-th/9308139; 
\\
``Some speculations about black hole entropy in string theory,''
arXiv: hep-th/9309145. 
\\
See also 
G. Veneziano, 
``A Stringy Nature Needs Just Two Constants'', 
Europhys. Lett. {\bf 2} (1986) 199; 
\\
M.~J.~Bowick, L.~Smolin and L.~C.~R.~Wijewardhana,
``Role Of String Excitations In The Last Stages Of Black Hole Evaporation,''
Phys.\ Rev.\ Lett.\  {\bf 56} (1986) 424; 
\\
M.~J.~Bowick, L.~Smolin and L.~C.~R.~Wijewardhana,
``Does String Theory Solve The Puzzles Of Black Hole Evaporation?,''
Gen.\ Rel.\ Grav.\  {\bf 19} (1987) 113. 

\bibitem{hp}
G.~T.~Horowitz and J.~Polchinski,
``A correspondence principle for black holes and strings,''
Phys.\ Rev.\ D {\bf 55} (1997) 6189, 
arXiv: hep-th/9612146.

\bibitem{fs}
W.~Fischler and L.~Susskind,
``Holography and cosmology,''
arXiv: hep-th/9806039.

\bibitem{bousso}
R.~Bousso,
``A Covariant Entropy Conjecture,''
JHEP {\bf 07} (1999) 004, 
arXiv: hep-th/9905177; 
\\
``Holography in general space-times,''
JHEP {\bf 06} (1999) 028, 
arXiv: hep-th/9906022; 
\\
``The holographic principle,''
Rev.\ Mod.\ Phys.\  {\bf 74} (2002) 825, 
arXiv: hep-th/0203101.

\bibitem{w}
See, for example, 
S. Weinberg,
``{\em Gravitation and cosmology : Principles and 
applications of the general theory of relativity}, 
John Wiley \& Sons, Inc. (1972). 

\bibitem{s2}
E.~Halyo, A.~Rajaraman and L.~Susskind,
``Braneless black holes,''
Phys.\ Lett.\ B {\bf 392} (1997) 319, 
arXiv: hep-th/9605112; 
\\
E.~Halyo, B.~Kol, A.~Rajaraman and L.~Susskind,
``Counting Schwarzschild and charged black holes,''
Phys.\ Lett.\ B {\bf 401} (1997) 15, 
arXiv: hep-th/9609075.

\bibitem{hprev}
G.~T.~Horowitz,
``Quantum states of black holes,''
arXiv: gr-qc/9704072; 
\\ 
A.~W.~Peet,
``The Bekenstein formula and string theory (N-brane theory),''
Class.\ Quant.\ Grav.\  {\bf 15} (1998) 3291, 
arXiv: hep-th/9712253. 

\bibitem{hp2}
G.~T.~Horowitz and J.~Polchinski,
``Self gravitating fundamental strings,''
Phys.\ Rev.\ D {\bf 57} (1998) 2557, 
arXiv: hep-th/9707170; 
\\
T.~Damour and G.~Veneziano,
``Self-gravitating fundamental strings and black holes,''
Nucl.\ Phys.\ B {\bf 568} (2000) 93, 
arXiv: hep-th/9907030.

\bibitem{rk}
S.~Kalyana Rama,
``Size of black holes through polymer scaling,''
Phys.\ Lett.\ B {\bf 424} (1998) 39, 
arXiv: hep-th/9710035;
\\
R.~R.~Khuri,
``Black holes and strings: The polymer link,''
Mod.\ Phys.\ Lett.\ A {\bf 13} (1998) 1407, 
arXiv: gr-qc/9803095; 
\\
``Self-gravitating strings and string/black hole correspondence,''
Phys.\ Lett.\ B {\bf 470} (1999) 73, 
arXiv: hep-th/9910122; 
\\
``Entropy and string / black hole correspondence,''
Nucl.\ Phys.\ B {\bf 588} (2000) 253, 
arXiv: hep-th/0006063.

\bibitem{bf}
T.~Banks and W.~Fischler,
``M-theory observables for cosmological space-times,''
arXiv: hep-th/0102077;
\\
``An holographic cosmology,''
arXiv: hep-th/0111142; 
\\
``Entropy of the stiffest stars,''
Class.\ Quant.\ Grav.\  {\bf 19} (2002) 4717, 
arXiv: hep-th/0206096; 
\\
``Black crunch,''
arXiv: hep-th/0212113;
\\
``Holographic cosmology 3.0,''
Phys.\ Scripta {\bf T117} (2005) 56, 
arXiv: hep-th/0310288; 
\\
``Holographic cosmology,''
arXiv: hep-th/0405200; 
\\
T.~Banks, W.~Fischler and L.~Mannelli,
``Microscopic quantum mechanics of the 
$p = \rho$ universe,''
Phys.\ Rev.\ D {\bf 71} (2005) 123514, 
arXiv: hep-th/0408076.
\\
See also 
M.~J.~Bowick and L.~C.~R.~Wijewardhana,
``Superstring Gravity And The Early Universe,''
Gen.\ Rel.\ Grav.\  {\bf 18} (1986) 59; 
\\
G.~Veneziano,
``A model for the big bounce,''
JCAP {\bf 03} (2004) 004, 
arXiv: hep-th/0312182.

\end{thebibliography}
\end{document}